\documentstyle[twoside,fleqn,espcrc2,epsfig]{article}
\title{Indications for Criticality at Zero Curvature
in a 4d Regge Model of Euclidean Quantum Gravity\thanks{This 
work was in part supported by the US Department of Energy under
contract DE-FG02-97ER41022.} }
\author{Wolfgang Beirl\address{Market Research Inc., Paradise Island, Bahamas}
and Bernd A. Berg\address{Department of Physics, The
Florida State University, Tallahassee, FL~32306, USA}$^,$\address{School 
of Computational Science and Information Technology,\\
~~The Florida State University, Tallahassee, FL 32306, USA} }
\begin{document}
\begin{abstract}
We re-examine the approach to four-dimensional Euclidean quantum gravity 
based on the Regge calculus. A cut-off on the link lengths is introduced 
and consequently the gravitational coupling and the cosmological constant become 
independent parameters. We determine the zero curvature, 
$\langle R\rangle =0$, line in the coupling constant plane by numerical 
simulations. When crossing this line we find a strong, probably first 
order, phase transition line with indications of a second order endpoint.
Beyond the endpoint the transition through the $\langle R\rangle =0$
line appears to be a crossover. Previous investigations, using the
Regge or the Dynamical Triangulation approach, dealt with a limit in
which the first order transition prevails.
\end{abstract}
\date{\today}
\maketitle

\section{Introduction}

Euclidean quantum gravity in four dimensions is formally defined by the 
path integral
\begin{equation} \label{Z1}
Z = \int Dg\, e^{ - S_{EH}[g] }\ .
\end{equation}
The Einstein-Hilbert action $S_{EH}[g]$ is a function of the 4-geometry 
$g$, $S_{EH}[g] = -L_P^{-2} R[g] + \Lambda V[g]$. The Planck length 
$L_P$ determines the gravitational coupling constant for the total 
curvature $R[g]$ and the total 4-volume $V[g]$ enters the action if 
the cosmological constant $\Lambda$ is non-zero.

In order to {\it define} the integral~(\ref{Z1}), we consider a lattice 
approximation, so that the path-integral becomes the limit of a 
series of well-defined and finite integrals. A direct route for such 
an approximation is provided by the Regge calculus~\cite{Regge}, which
is suitable for simulations~\cite{Berg,Hamber}. One 
considers a simplicial lattice of fixed connectivity, which subdivides 
a 4-manifold of given topology in $N_4$ 4-simplices, $N_3$ tetraeder, 
$N_2$ triangles, $N_1$ links and $N_0$ vertices. As we assign a length 
to each link, 
a 4-geometry is defined in the Euclidean sector, if the triangle 
inequalities are fulfilled for every (sub-)simplex. In the following 
we will use the squared lengths $q_l$ associated with each link $l$ as 
the basic variables and a given set of such variables defines
a 4-geometry $[q]$ for a given simplicial lattice. Variation of the 
variables $q_l$ allows us to approximate the path integral (\ref{Z1}). 
One can calculate the Regge action $R[q] = \sum_t A_t \delta_t$ as
a sum over all triangles $t$ with area $A_t$ multiplied by their
associated deficit angles $\delta_t$. 
The 4-volume is calculated as a sum over the volume 
of all 4-simplices $s$, $V[q] = \sum_s V_s$. We then approximate the 
gravitational action by $S_{EH} \rightarrow S[q] = -L_P^{-2} R[q] + 
\Lambda V[q]$. It has been demonstrated how the geometrical variables 
of the Regge calculus converge towards the corresponding variables of 
General Relativity in the classical continuum 
limit~\cite{Regge,classical_limit}. It requires an upper bound on the 
link lengths, $q_l < Q$, and a limit $F$ on the 'fatness' 
$\phi_s = V_s/\max(q_l^2)$ of each 4-simplex, $\phi_s > F$. The 
continuum limit is then reached for $N_0 \to \infty$, 
$Q \to 0$ with $F > 0$ (Cheeger et al.~\cite{classical_limit}).

In order to replace the path integral (\ref{Z1}) by a finite
integral on the simplicial lattice, 
\begin{equation} \label{Z2}
Z = \int Dq\, e^{ - S[q] }.
\end{equation}
we need to define the integration measure. A simple choice is the 
uniform measure
\begin{equation} \label{Z3}
 Dq = \int_0^Q \prod_l dq_l \Theta_F[q] 
\end{equation}
where the function $\Theta_F$ equals $1$ when all triangle inequalities
are fulfilled and $\phi_s > F$ for all 4-simplices; it is $0$ otherwise.
The integration limit $Q$ is necessary 
for a well-defined integral, but it has no physical meaning in itself. 
Indeed, simple re-scaling $q_l \to Q q_l$ formally removes it and leads 
to the integral
\begin{equation} \label{Z4}
 Z(\beta,\lambda) = \int_0^1 \prod_l dq_l \Theta_F[q] 
e^{ \beta R[q] - \lambda V[q] }
\end{equation}
where $\beta = Q L_P^{-2}$ and $\lambda = Q^2 \Lambda$. 

In previous numerical investigations~\cite{HamberJanke,Vienna} the upper 
limit $Q$ was infinite which corresponds to the limit $\beta \to \infty$ 
and $\lambda \to \infty$ in the integral (\ref{Z4}).

In the following we will numerically evaluate the integral (\ref{Z4})  
in order to determine the phase structure by varying the coupling parameters
$\beta$ and $\lambda$. Our results suggest a physically interesting continuum
limit for finite values of the coupling parameters.      

\section{Results}

The numerical computations have been performed for the integral
\begin{equation} \label{Z5}
 Z(\beta,\lambda) = \int_0^c \prod_l dq_l \Theta_F[q] 
 e^{ \beta R[q] - \lambda V[q] }
\end{equation}
with $c = 10$ instead of  (\ref{Z4}) with $c= 1$. This means that 
the coupling constants and expectation values need to be adjusted 
accordingly, e.g. $\beta \to c\beta$, $\lambda \to c^2\lambda$ etc.
We set $F = 10^{-6}$ but our results seem to be independent of the
specifc value for a relatively wide range $F = 10^{-8}$ to $10^{-5}$.   
We used a hypercubic regular triangulation of the 4-torus~\cite{Berg,Hamber}, 
with $N_0= 4^4, 4^3 \times 10, 6^4$ and $8^4$ vertices.

\begin{figure}[ht] \vspace{-2mm} \begin{center}
\epsfig{figure=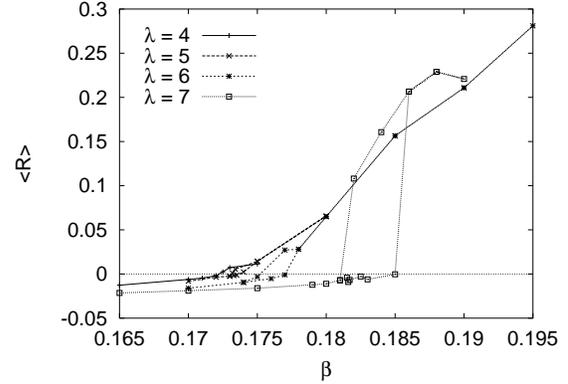,width=\columnwidth} \vspace{-15mm}
\caption{ Hysteresis loops. }
\label{fig_1b} \end{center} \vspace{-15mm}
\end{figure}

\begin{figure}[ht] \vspace{-2mm} \begin{center}
\epsfig{figure=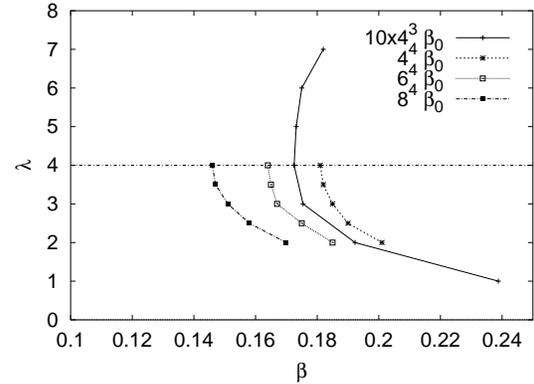,width=\columnwidth} \vspace{-15mm}
\caption{ Function $\beta_0(\lambda)$ for which $<R>=0$ holds. }
\label{fig_2} \end{center} \vspace{-9mm}
\end{figure}

From simulations with $\lambda = 1, 2, 3, 4$ and varying $\beta$ we
find expectation values $<R>$, which are (almost) zero for 
$\beta_0( \lambda )$ values; $\beta_0$ initially decreases for increasing 
$\lambda$. Then, as depicted in figure~\ref{fig_1b}, we find that 
$\beta_0$ increases for $\lambda = 4, 5, 6, 7$. Furthermore, the 
transition from $<R> < 0$ to $<R> > 0$ is no longer defined clearly, 
because we have to deal with a significant hysteresis for large 
$\lambda$. The line $\beta_0( \lambda )$ is depicted in 
figure~\ref{fig_2} for different lattice size.
The curve $\beta_0( \lambda )$ is interesting, since a vanishing 
curvature is a desired ingredient for a physical continuum limit.

\begin{figure}[ht] \vspace{-2mm} \begin{center}
\epsfig{figure=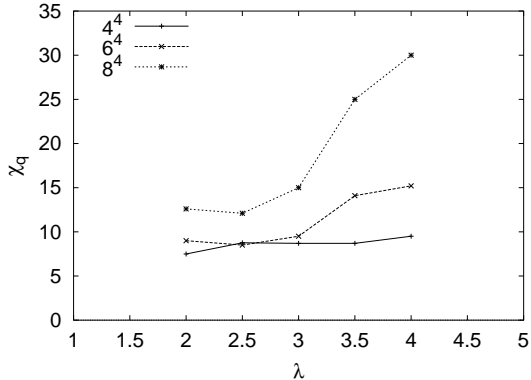,width=\columnwidth} \vspace{-15mm}
\caption{ Susceptibility $\chi_q$ as function of $\lambda$ at $\beta_0$ for 
increasing lattice size. }
\label{fig_3a} \end{center} \vspace{-15mm}
\end{figure}

As we approach the turning point $\lambda_c \sim 4.0$ from smaller 
$\lambda$, the suszeptibility $\chi_q\ =\ N_1^{-1}\ 
(<\overline{q}^2>-<\overline{q}>^2)$, with 
$\overline{q} = N_1^{-1} \sum_l q_l$, increases significantly; this 
increase is more pronounced for larger lattice size as depicted in 
figure~\ref{fig_3a}.  

\begin{figure}[ht] \vspace{-2mm} \begin{center}
\epsfig{figure=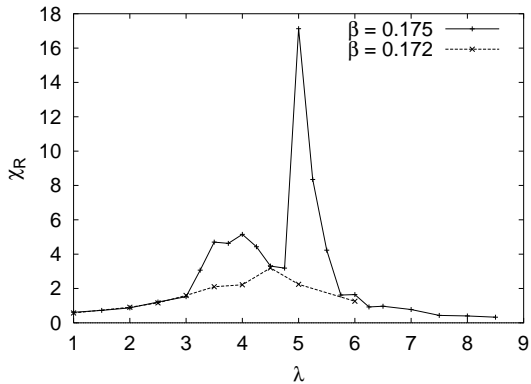,width=\columnwidth} \vspace{-15mm}
\caption{ Susceptibility $\chi_R$ as function of $\lambda$ for 
two $\beta$ values. }
\label{fig_3b} \end{center} \vspace{-9mm}
\end{figure}

Next, we performed calculations varying $\lambda$ at $\beta=0.175$
and $0.172$ for the lattice size $4^3 \times 10$.
As expected, we find that the curvature $<R>$ increases initially,
exhibiting a transition from negative to positive values, but when 
$\lambda$ increases further a second transition back to negative 
values is seen.
We depict $\chi_R = N_2^{-1}( <\overline{R}^2> - <\overline{R}>^2 )$,
measuring curvature fluctuations, in figure~\ref{fig_3b}; for
$\beta=0.175$ one can distinguish two peaks corresponding to the
two transitions. These peaks merge into one at $\beta = 0.172$.

We also measured the 2-point correlation function $K(d)=<q_0 q_d>-<q_0>^2$ 
along the elongated direction on the $4^3 \times 10$ lattice for
$\beta_0( \lambda )$ and found 
that the function $K(d)$ is significantly different from zero for $d>1$. 
Negative values alternate with positive values, which indicates a 
non-trivial interaction. The correlations increase for $\lambda\to\lambda_c$.

Although we are not able to provide a complete finite size 
scaling analysis, our data suggest the following: In the region
$\lambda > \lambda_c$, with $\lambda\sim 4.0$, one deals with a 
strong phase transition, which is likely of first order. In the 
region $\lambda < \lambda_c$ we are currently unable to distinguish
between a cross-over transition or a higher-order phase transition.
However the end-point $\lambda_c$ itself is a natural candidate for 
a second-order phase transition. The increase of susceptiblity $\chi_q$ 
and the link correlations $K(d)$ approaching $\lambda_c$ from smaller 
values confirm this picture.  

\section{Summary and Conclusions}

Our simulations differ from those of the previous literature by
introducing the limit $Q$ in the functional integral~(\ref{Z3}).
While a simple re-scaling removes $Q$ formally from the partition 
function~(\ref{Z4}), our model remains more general than those 
simulated in the literature, which are recovered in our limit of
large $\lambda$. Remarkably, the new model appears to have
a second order fixed point in the $\lambda-\beta$ plane, which 
deserves further study.


\begin{thebibliography}{99}

\bibitem{Regge} T. Regge, Nuovo Cimento {\bf 9} (1961) 558.

\bibitem{Berg} B.A. Berg, Phys. Rev. Lett. {\bf 55} (1985) 904.

\bibitem{Hamber} H.W. Hamber and R.M. Williams, Phys. Lett. B
{\bf 157} (1985) 368. 

\bibitem{classical_limit} R. Friedberg and T.D. Lee, Nucl. Phys. B
{\bf 242} (1984) 145; G. Feinberg, R. Friedberg, T.D. Lee and H.C. Ren,
Nucl. Phys. B {\bf 245} (1984) 343; H. Cheeger, W. M\"uller and 
R. Schrader, Commun. Math. Phys. {\bf 92} (1984) 405.

\bibitem{HamberJanke} M. Gross and H. Hamber, Nucl. Phys. B {\bf 364} 
(1991) 703; C. Holm and W. Janke, Phys. Lett. B {\bf 335} (1994) 143.

\bibitem{Vienna} W. Beirl, P. Homolka, B. Krishnan, M. Markum and
J. Riedler, Nucl. Phys. B (Proc. Suppl.) 42, 710 (1995); J. Riedler,
W. Beirl, E. Bittner, A. Hauke, P. Homolka and H. Markum Class. 
Quant. Grav. 16, 1163 (1999).

\end{thebibliography}
\end{document}